\documentclass[12pt,letterpaper]{article}
\usepackage{amsmath,amssymb,epsfig}
\numberwithin{equation}{section}
\usepackage{cite}
\newcommand{\be}{\begin{equation}}
\newcommand{\bea}{\begin{eqnarray}}
\newcommand{\eea}{\end{eqnarray}}
\newcommand{\ba}{\begin{array}}
\newcommand{\ea}{\end{array}}
\newcommand{\ee}{\end{equation}}

\expandafter\ifx\csname mathbbm\endcsname\relax

\else

\fi \textheight 22cm \textwidth 15cm \topmargin 1mm \oddsidemargin
5mm \evensidemargin 5mm

\begin{document}

\begin{titlepage}
\hfill
\vbox{
    \halign{#\hfil         \cr
           IPM/P-2006/038 \cr
                      } % end of \halign
      }  % end of \vbox
\vspace*{20mm}
\begin{center}
{\Large {\bf New Attractors, Entropy Function and Black Hole
Partition Function}\\
}

\vspace*{15mm}
\vspace*{1mm}
{Mohsen Alishahiha\footnote{alishah@theory.ipm.ac.ir} and
Hajar Ebrahim\footnote{ebrahim@theory.ipm.ac.ir}}
\vspace*{1cm}

{\it  Institute for Studies in Theoretical Physics
and Mathematics (IPM)\\
P.O. Box 19395-5531, Tehran, Iran \\ \vspace{3mm}}

\vspace*{1cm}
%%\maketitle
\end{center}

\begin{abstract}
By making use of the entropy function formalism we study the
generalized attractor equations in the four dimensional ${\cal
N}=2$ supergravity in presence of higher order corrections. This
result might be used to understand a possible ensemble one could
associate to an extremal black hole.
%Using the generality
%and simplicity of this formalism we establish a duality between a
%four gravitational theory on $AdS_2\times S^2$ background and the
%extremal black hole of the theory whose near horizon geometry is
%fixed by the $AdS_2$ background. In this sense the attractor
%mechanism plays the role of decoupling limit in the context of
%AdS/CFT correspondence.
\end{abstract}
\end{titlepage}

\section{Introduction}

Recently the black hole attractor mechanism has attracted a lot of attention.
This is mainly because of the recent developments in the connection between
the partition
function of four dimensional BPS black holes and partition function of
topological strings
\cite{Ooguri:2004zv}.

According to the attractor mechanism the values of the moduli
scalar fields at the horizon are entirely determined by the
charges of the black hole regardless of their asymptotic values.
Originally this special behavior has been discovered in the
context of BPS extremal black holes in four dimensional ${\cal
N}=2$ supergravity with unbroken supersymmetry in
\cite{{Ferrara:1995ih},{Strominger:1996kf},{Ferrara:1996dd},{Ferrara:1996um}}.
Later on it was shown  that the attractor mechanism can also work
for non-BPS extremal black holes
\cite{{Ferrara:1997tw},{Gibbons:1996af},{Tripathy:2005qp}}. In
particular new algebraic attractor equations describing both BPS
and non-BPS solutions have been introduced in
\cite{{Kallosh:2005ax}} (for further discussions see
\cite{{Giryavets:2005nf},{Dall'Agata:2006nr},{Kallosh:2006bt},
{Bellucci:2006ew}}).

More recently attractor mechanism has been studied in
non-supersymmetric extremal black holes \cite{{Goldstein:2005hq}}
(see also \cite{Chandrasekhar:2006kx}). In fact a similar
structure to supersymmetric extremal black holes appears in
non-supersymmetric cases. Namely the true attractive points
correspond to the critical points of the black hole effective
potential which make the potential minimum. Moreover the entropy
of these non-supersymmetric extremal black holes is given by the
value of the effective potential at the extremum and therefore due
to the attractor behavior, it is given by the charges of the black
hole.

On the other hand a general method for computing the entropy of
spherically symmetric extremal black holes in a theory of gravity
coupled to gauge fields and scalar fields has been developed in
\cite{Sen:2005wa}. In this method one can obtain the entropy of
the extremal black hole just by using its near horizon field
configuration, assuming the existence of the full black hole
solution. To be precise let us consider an extremal
$d$-dimensional black hole whose near horizon geometry is
$AdS_2\times S^{d-2}$ and carries electric and magnetic charges.
There are also several scalar fields in the theory. One then
defines the entropy function as the Legendre transform of the
Lagrangian density integrated over angular variables, with respect
to the value of the electric field strength at the horizon.
Extremizing the entropy function with respect to near horizon
variables will result in a set of algebraic equations for these
parameters. The entropy of these black holes is given by the value
of the entropy function evaluated at the extremum \footnote{We
note that a similar function has also been found in the
supersymmetric case in \cite{{Behrndt:1996jn},{Mohaupt:2005jd}}.
Although in the supersymmetric case it is not called entropy
function, it has precisely the same properties.}.

It is shown that the entropy function is actually proportional to
the effective potential of non-supersymmetric extremal black holes
\cite{Alishahiha:2006ke}. In this sense the entropy function is
given directly in terms of the prepotential in the supersymmetric
case. We will come back to this point in section 4.

It is worth noting that due to the algebraic nature of the
equations, higher derivative corrections to the action and entropy
can be obtained more simply in this formalism. Indeed this method
has been used to compute corrections to the entropy of the
different extremal black holes in
\cite{{Sen:2005iz},{Prester:2005qs},{Sinha:2006yy},{Chandrasekhar:2006zw},
{Ghodsi:2006cd}}.

Although the entropy function has mainly been used in
non-supersymmetric theories, this formalism has also been applied
to ${\cal N}=2$ supergravity theories in \cite{Sahoo:2006rp} where
the authors have shown that the BPS attractor equations can be
obtained by extremizing the  entropy function with respect to the
black hole charges. It is one of the aims of this paper to study
the generalized attractor equations using the entropy function
which can be applied to supersymmetric as well as
non-supersymmetric theories. Being simple, this formalism can be
used to find the attractor equations in the presence of higher
order corrections. The supersymmetric attractor equations in the
presence of higher order corrections have been studied in
\cite{{LopesCardoso:1998wt}}.

Having had the attractor equations which come out from the
equations of motion, one may ask if the entropy function mechanism
is just a technical method. In other words we would like to
understand the physical interpretation (if any) of the entropy
function which is essentially equivalent to the Wald formula for
the entropy. In order to address this question we will follow the
recent works on connection between topological string theories and
black hole partition function \cite{Ooguri:2004zv} where the
authors have proposed a mixed ensemble for the extremal black hole
of four dimensional supergravity obtained by compactification of
type IIA on Calabi-Yau 3-fold. In particular we would like to
compare the entropy function formalism with the structure used in
\cite{Ooguri:2004zv} (see also
\cite{{LopesCardoso:2006bg},{Dabholkar:2004yr},{Dabholkar:2005by},
{Dabholkar:2005dt},{Parvizi:2005aa}}).

This paper is organized as follows. In section 2 we rederive the
generalized attractor equations in ${\cal N}=2$ supergravity
theories by extremizing the effective potential. In section 3 we
show how the generalized attractor equations can be obtained in
entropy function formalism where higher order corrections can also
be taken into account. In section 4 we will study the partition
function of the extremal black holes in the context of entropy
function formalism following OSV conjecture. The last section is
devoted to discussions.

\section{Generalized attractor equations}

In this section we rederive generalized attractor equations in
${\cal N}=2$ four dimensional supergravity coupled to $n$ vector
multiplets. These equations have recently been studied in
\cite{{Giryavets:2005nf},{Dall'Agata:2006nr},{Kallosh:2006bt},
{Bellucci:2006ew}}. Using these equations one can study BPS and
non-BPS attractive points in the same way. The aim of this section
is to present a direct derivation of these generalized attractor
equations by minimizing the leading order effective potential.

Consider ${\cal N}=2$ four dimensional supergravity coupled to $n$
vector multiplets. To study these theories it is useful to work
within the framework of the special geometry. A special K\"ahler
manifold can be constructed by a $2n+2$ dimensional flat
symplectic bundle over a K\"ahler manifold with a symplectic
section defined by \be\Pi=(L^I,M_I),\;\;\;I=0,..,n, \label{pi} \ee
subject to a constraint $i({\bar L}^IM_I-L^I\bar{M}_I)=1$. $L^I$
and $M_I$ depend on scalar fields $t$ and ${\bar t}$  which
parameterize  the moduli space. They are also covariantly
holomorphic which means $D_{\bar i}\Pi=(\partial_{\bar
i}-\frac{1}{2}K_{\bar i})\Pi=0$. Here $K$ is the K\"ahler
potential.

Introducing a symplectic charge $(q_I,p^I)$ one can define a
covariantly holomorphic central charge as \be
Z(t,\bar{t},p,q)\equiv (L^Iq_I-M_Ip^I), \ee which satisfies
$D_{\bar i}Z=D_i{\bar Z}=0$. From four dimensional supergravity
point of view one may identify this with the charge of the
graviphoton. On the other hand one may identify the $(q_I,p^I)$
with the charges of a black hole solution in this four dimensional
supergravity whose effective potential is given by \be
V_{eff}=\left|Z\right|^2+\left|D_iZ\right|^2,\ee which is
symplectic invariant. It is known that the attractor equations can
be obtained by extremizing this potential
\cite{{Ferrara:1997tw},{Denef:2000nb}}.

The extremization of the effective potential will give the
following condition \be
2(D_iZ)\bar{Z}+G^{j\bar{k}}D_{i}D_{j}Z\bar{D}_{\bar{k}}\bar{Z}=0,
\label{veff} \ee which can be solved to find the attractor points.
One of its solutions is given by \be
D_iZ=\bar{D}_{\bar{i}}\bar{Z}=0,\ee which leads to the
supersymmetric attractor equations as follows \be
p^I=i(\bar{Z}L^I-Z\bar{L}^I),\;\;\;q_I=i(\bar{Z}M_I-Z\bar{M}_I).
\ee
One may also relax the supersymmetric condition $D_iZ=0$ and
look for a general solution of equation (\ref{veff}) which could
lead to non-supersymmetric equations as well.
To do this let us start with the conjugate form of the equation
(\ref{veff}). Using the definition of $Z$ the first term in the
equation (\ref{veff}) reads \be
2Z\bar{D}_{\bar{i}}\bar{Z}=2Z(D_{\bar{i}}\bar{L}^Iq_I-D_{\bar{i}}\bar{M}_Ip^I).
\ee One can replace $D_{\bar{i}}\bar{M}_I$ with
${\cal{N}}_{IJ}D_{\bar{i}}\bar{L}^J$, where ${\cal N}_{IJ}$ is a
complex symmetric $(n+1)\times (n+1)$ matrix such that
$D_{\bar{i}}\bar{M}_I={\cal{N}}_{IJ}D_{\bar{i}}\bar{L}^J$.  Now
contracting both sides of the equation with $G^{i\bar{i}}D_iL^K$
and  using the identity \be
D_iL^IG^{i\bar{i}}D_{\bar{i}}\bar{L}^J=-\frac{1}{2}{\rm
Im}({\cal{N}}^{-1})^{IJ}-\bar{L}^IL^J, \ee we will obtain the
following relation for the charges $p^I$ \bea
p^I=i\bigg{[}-2Z\bar{L}^I+\frac{G^{i\bar{j}}G^{l\bar{k}}\bar{D}_{\bar{j}}
\bar{D}_{\bar{k}}\bar{Z}D_lZD_iL^I}{Z}
-\left[{\rm Im}({\cal N}^{-1})q+({\rm Re}{\cal N}{\rm Im}{\cal
N}^{-1})p\right]^I\bigg{]}. \eea To get the other charges $q_I$ we
first multiply the above equation by $(\bar{{\cal N}}^{-1})^{IL}$
and then taking into account that ${\cal N}_{IJ}({\cal
N}^{-1})^{JL}=\delta_I^L$, we arrive at \bea
q_I&=&i\bigg{[}-2Z\bar{M}_I+\frac{G^{i\bar{j}}G^{l\bar{k}}\bar{D}_{\bar{j}}
\bar{D}_{\bar{k}}\bar{Z}D_lZD_iM_I}{Z}\cr
&+&\left[({\rm Re}{\cal N}{\rm Im}{\cal N}^{-1}{\rm Re}{\cal
N}+{\rm Im}{\cal N})p -({\rm Im}{\cal N}^{-1}{\rm Re}{\cal
N})q\right]_I\bigg{]}. \eea Now taking the imaginary parts of the
above relations we can obtain the generalized attractor equations
which are \bea q_I&=&2{\rm Im}
\bigg{[}Z\bar{M}_I-\frac{G^{i\bar{j}}G^{l\bar{k}}\bar{D}_{\bar{j}}
\bar{D}_{\bar{k}}
\bar{Z}D_lZ}{2Z}D_iM_I\bigg{]},\cr &&\cr p^I&=&2{\rm Im}
\bigg{[}Z\bar{L}^I-\frac{G^{i\bar{j}}G^{l\bar{k}}\bar{D}_{\bar{j}}
\bar{D}_{\bar{k}}
\bar{Z}D_lZ}{2Z}D_iL^I\bigg{]}. \label{matt} \eea Here we assume
that $Z\neq 0$. This is the general form of the attractor
equations in leading order which are valid for both supersymmetric
and non-supersymmetric cases. In particular setting $D_iZ=0$ one
gets the supersymmetric equations. They have exactly the same form
as those in \cite{Kallosh:2006bt}.

\section{Generalized attractor equations from entropy function }

In this section we shall study the attractor equations using the
entropy function formalism \cite{Sen:2005wa}. It is important to
note that the attractor equations presented in the previous
section are valid only in leading order. In fact taking into
account higher order corrections to the action one needs to
minimize the corrected effective potential. The supersymmetric
attractor equations in presence of higher order corrections have
been studied in \cite{{LopesCardoso:1998wt}}.

We note, however, that using the entropy function one may easily handle the
higher order
corrections in the same way as the leading order both for supersymmetric
and non-supersymmetric cases.
Actually it is the aim of this section to derive the
generalized attractor equations for ${\cal N}=2$ supergravity in four
dimensions when
the higher order corrections are also taken into account.

\subsection{General formalism}

Let us first review the minimum ingredients we need to write  the
off-shell form of ${\cal N}=2$ supergravity action in four
dimensions (see for example
\cite{{deWit:1979ug},{deWit:1980tn},{deWit:1983rz}}). In what
follows we use the notation of \cite{Mohaupt:2000mj}. To study
this theory it is useful to start with superconformal theory and
then we can fix the gauge to get the supergravity theory we are
interested in. The representation of the corresponding
superconformal algebra contains Weyl, vector and non-linear
multiplets.

Since we are interested in the off-shell representation, these
multiplets contain dynamical bosonic fields, the corresponding
fermionic superpartners and non-dynamical fields. The dynamical
bosonic fields of the theory are $(N+1)$ complex scalars $X^I$
with $0\leq I\leq N$, metric $G_{\mu\nu}$ and $(N+1)$ gauge fields
$A_\mu^I$. The non-dynamical fields of the multiplets are a
complex anti-self-dual antisymmetric tensor field $T_{\mu\nu}^-$,
a real scalar field $D$, a $U(1)$ gauge field ${\cal A}_\mu$, an
$SU(2)$ gauge field ${\cal V}^i_{j\mu}$, a vector field $V_\mu$, a
set of $SU(2)$ triplet scalar fields $Y_{ij}^I$, an $SU(2)$
triplet scalar field $M_{ij}$ and scalar field $\Phi_i^\alpha$
which transform as a fundamental of both the gauge $SU(2)$ and
global $SU(2)$ symmetries. Here $i,j=1,2$ are $SU(2)$ indices
which are raised and lowered by the anti-symmetric tensor
$\epsilon^{ij}$ and $\epsilon_{ij}$. There are also fermionic
fields which are not presented here.

In this formulation the action involving  these fields can be
written in terms of the prepotential $F(X^I, \hat{A})$, which is a
homogeneous function of the complex scalars $X^I$ and the
composite auxiliary field ${\hat A}=T^{-\mu\nu}T^-_{\mu\nu}$ such
that \be F(\lambda X^I,\lambda^2{\hat A})=\lambda^2F(X^I,{\hat
A}). \ee In terms of the prepotential, defining \be
F_I=\frac{\partial F}{\partial X^I},\;\;\;F_{\hat
A}=\frac{\partial F}{\partial {\hat A}},
\;\;\;F_{IJ}=\frac{\partial^2 F}{\partial X^I\partial X^J},\;\;\;
F_{I{\hat A}}=\frac{\partial^2 F}{\partial X^I\partial {\hat
A}},\;\;\; F_{{\hat A}{\hat A}}=\frac{\partial^2 F}{\partial {\hat
A}\partial {\hat A}}, \ee the bosonic part of the Lagrangian is
given by ( see equation (3.111) of \cite{Mohaupt:2000mj}) \bea
8\pi{\cal L}&=&-\frac{i}{2}(X^I{\bar F}_I-{\bar X}^IF_I)R+\bigg{[}
i(\partial_\mu F_I+i{\cal A}_\mu F_I)(\partial^\mu {\bar
X}^I-i{\cal A}^\mu {\bar X}^I) +\frac{i}{32}{\bar F}{\hat A}\cr
&&\cr &+&\frac{i}{4}F_{IJ}A^I_{\mu\nu}A^{J\mu\nu}+\frac{i}{8}{\bar
F}_IA^I_{\mu\nu}T^{-\mu\nu} +\frac{i}{2}{\hat
F}^-_{\mu\nu}F_{I{\hat A}}A^{I\mu\nu}
-\frac{i}{8}F_{IJ}Y^I_{ij}Y^{Jij} +\frac{i}{2}F_{\hat A}{\hat C}
\cr &&\cr &-& \frac{i}{8}F_{{\hat A}{\hat A}}({\hat B}_{ij}{\hat
B}^{ij} -2{\hat F}^{-}_{\mu\nu}{\hat F}^{-\mu\nu})
-\frac{i}{4}{\hat B}_{ij}F_{I{\hat A}}Y^{Iij}+h.c.\bigg{]}\\
&-&i(X^I{\bar F}_I-{\bar X}^IF_I)\left(\nabla^\mu
V_\mu-\frac{1}{2}V^\mu V_\mu-\frac{1}{4}|M_{ij}|^2
+|\partial_\mu\Phi^{\alpha}_i+\frac{1}{2}{\cal
V}^k_{i\mu}\Phi^\alpha_k|^2\right),\nonumber
\eea
where
$A^I_{\mu\nu}=F^{I-}_{\mu\nu}-\frac{1}{4}{\bar X}^IT^-_{\mu\nu}$
with $F^{I-}_{\mu\nu}=\frac{1}{2}(F^I_{\mu\nu}- i\; {}^\ast
F^{I}_{\mu\nu})$. As we will see this particular combination plays
an important role. Note that the fields are  subject to the
constraint \be \nabla^\mu V_\mu-\frac{1}{2}V^\mu
V_\mu-\frac{1}{4}|M_{ij}|^2
+|\partial_\mu\Phi^{\alpha}_i+\frac{1}{2}{\cal
V}^k_{i\mu}\Phi^\alpha_k|^2=D-\frac{1}{3}R. \ee
For more details and also the definition of
other components which we have used here see
\cite{Mohaupt:2000mj}.

Let us now consider an extremal black hole in this supergravity
theory with near horizon geometry of the form \cite{Sahoo:2006rp}
\bea
ds^2&=&v_1(-r^2dt^2+\frac{dr^2}{r^2})+v_2(d\theta^2+\sin^2\theta
d\phi^2),\cr &&\cr
F^I_{rt}&=&e^I,\;\;\;\;F^{I}_{\theta\phi}=p^I\sin\theta,\;\;\;\;X^I=x^I,\;\;
\;\;
T^-_{rt}=v_1 \omega. \label{ansatz}\eea The other fields are given
by \be {\cal A}_\mu=0,\;\;\;{\cal
V}^i_{j\mu}=0,\;\;\;V_\mu=0,\;\;\;M_{ij}=0,\;\;\;Y^I_{ij}=0,\;\;\;
\Phi^\alpha_i=\delta^\alpha_i,\;\;\;D-\frac{1}{3}R=0. \ee It is
easy to see that this is a consistent truncation. Note that for
the ansatz we are considering we have \be
\hat{A}=-4\omega^2,\;\;\;\;A^{I}_{\mu\nu}\equiv
v_1A^I=e^I-i\frac{v_1}{v_2}p^I-\frac{1}{2}{\bar x}^Iv_1\omega.
\label{A} \ee Following \cite{Sen:2005wa} the entropy function is
defined as \be {\cal
E}(v_1,v_2,\omega,x^I,e^I,q_I,p^I)=2\pi\left(-\frac{1}{2}
q_Ie^I-\int d\theta d\phi \sqrt{-G}{\cal L}\right) \ee which for
our ansatz it reads \bea {\cal E}&=&-\pi q_I e^I-\pi
v_1v_2\bigg{\{}i(v_1^{-1}-v_2^{-1})(x^I{\bar F}_I-{\bar
x}^IF_I)+\frac{i}{8}({\bar \omega}^2 F-\omega^2 {\bar F})\cr &&\cr
&-&\frac{i}{4}F_{IJ}A^IA^J+\frac{i}{4}{\bar F}_{IJ}{\bar A}^I{\bar
A}^J -\frac{i}{4}\omega {\bar F}_{I}A^I+\frac{i}{4}{\bar \omega} {
F}_{I}{\bar A}^I\\
&+& 8i\omega{\bar \omega}(-v_1^{-1}-v_2^{-1}+\frac{1}{8}
\omega{\bar \omega})(F_{\hat{A}}-{\bar F}_{\hat{A}})
+64i(v_1^{-1}-v_2^{-1})^2(F_{\hat{A}}-{\bar
F}_{\hat{A}})\bigg{\}}.\nonumber
\eea
In this framework the equations of motion can be obtained by extremizing the
entropy function
{\it i.e.} \be \frac{\partial {\cal E}}{\partial
v_i}=0,\;\;\;\;\;\;\frac{\partial {\cal E}}{\partial x^I}=0,
\;\;\;\;\;\;\frac{\partial {\cal E}}{\partial
\omega}=0,\;\;\;\;\;\; \frac{\partial {\cal E}}{\partial e^I}=0.
\ee The entropy function defined here is invariant under local
scale transformation \be x^I\rightarrow \lambda
x^I,\;\;\;v_i\rightarrow
\lambda^{-1}\bar{\lambda}^{-1}v_i,\;\;\;e^I\rightarrow e^I,\;\;\;
\omega\rightarrow\lambda\omega,\;\;\;q_I\rightarrow
q_I,\;\;\;p^I\rightarrow p^I. \ee
This is related to the conformal
symmetry of the ${\cal N}=2$ supergravity theory action. In
special geometry one can fix this symmetry using the
symplectic constraint on $(L^I,M_I)$ that is
$i(\bar{L}^IM_I-L^I\bar{M}_I)=1$.
In principle one should fix the
gauge, though, it is more convenient to work with gauge invariant
action. Later on we will fix the gauge in the level of equations
of motion. We note, however, that the gauge can be fixed in
several ways. In particular, following \cite{Sahoo:2006rp}, the
scaling symmetry in entropy function can be eliminated by imposing
the condition
\be \omega={\rm constant},\label{gauge}
\ee
on the equations of motion.

It is also worth noting that if we eliminate $e^I$, using the
equations of motion for $e^I$, $\frac{\partial {\cal E}}{\partial
e^I}=0$, one can see that entropy function is invariant under
symplectic transformation which acts on $(p^I,q_I)$ and
$(X^I,F_I)$ as follows: \be \left(\begin{array}{c}\check X^I \\
\check F_J\end{array}\right) = \left(\begin{array}{cc}U^I_{~K} &
Z^{IL} \cr W_{JK} & V_J^{~L}\end{array}\right)
\left(\begin{array}{c} X^K\cr F_L\end{array}\right),  \qquad
\left(\begin{array}{c}\check p^{I} \cr \check
q_{J}\end{array}\right) = \left(\begin{array}{cc}U^I_{~K} & Z^{IL}
\cr W_{JK} & V_J^{~L}\end{array}\right) \left(\begin{array}{c}p^{
K}\cr q_{L}\end{array}\right), \ee where $U$, $Z$, $W$ and $V$ are
each $(N+1)\times (N+1)$ matrices satisfying \be
U^TW-W^TU=0,\;\;\;Z^TV-V^TZ=0,\;\;\;U^TV-W^TZ=1.\ee This
symplectic invariance keeps other parameters unchanged. It should
be mentioned that in entropy function formalism, the set
$(X^I,F_I)$ plays the role of $(L^I,M_I)$ in special geometry.

Now we have all the ingredients we need to write the most general
form of the attractor equations using the entropy function
formalism. In fact the main purpose is to obtain the value of the
scalars or moduli fields at the horizon in terms of the electric
and magnetic charges of the black hole. Therefore as the first
step one needs to extremize the entropy function with respect to
$e^I$. Doing so, one gets \be
q_I=i\frac{v_2}{4}\bigg{[}(\omega\bar{F}_I-\bar{\omega}F_I)+2(F_{IJ}A^J-
\bar{F}_{IJ}\bar{A}^J)\bigg{]}. \label{q} \ee On the other hand
taking the real and imaginary parts of $A^I$ one finds \be
e^I=\frac{v_1}{4}\bigg{[}(\bar{\omega}x^I+\omega {\bar
x}^I)+2(A^I+\bar{A}^I) \bigg{]}, \label{e} \ee and \be
p^I=i\frac{v_2}{4}\bigg{[}(\omega\bar{x}^I-\bar{\omega}x^I)+2(A^I-\bar{A}^I)
\bigg{]}.\label{p} \ee The equations (\ref{q}) and (\ref{p}) are
actually the generalized attractor equations for ${\cal N}=2$
supergravity in four dimensions where the higher order corrections
have also been taken into account. In fact these equations should
be compared to those in (\ref{matt}) which are the generalized
attractor equations in leading order.

These equations can be applied to both BPS and non-BPS black hole
solutions. Actually the supersymmetric and non-supersymmetric
black holes correspond to the solutions with $A^I=0$ and $A^I\neq
0$, respectively. This is very similar to what we have in the
special geometry framework of these theories where supersymmetric
and non-supersymmetric solutions or attractor equations correspond
to $DZ=\bar{D}\bar{Z}=0$ and $DZ\neq 0,\;\bar{D}\bar{Z}\neq 0$,
respectively.

To understand the relation between supersymmetry and vanishing
$A^I$ better, it is useful to look at the supersymmetry
transformation of the spinor fields in the vector multiplet. In
particular consider the variation of gaugini under supersymmetry
transformation generated by $Q$ with the parameters $\epsilon_i$
in the notation of \cite{Mohaupt:2000mj} \be \delta
\Omega^I_i=2\gamma^\mu D_\mu
X^I\;\epsilon_i+\frac{1}{2}\epsilon_{ij} \gamma^\mu\gamma^\nu{\cal
F}_{\mu\nu}^{I-}\epsilon^j+ Y^I_{ij}\epsilon^j+2X^I\eta_i, \ee
where $D$ is the covariant derivative with respect to all
superconformal transformations and \be {\cal
F}^{I-}_{\mu\nu}=F^{I-}_{\mu\nu}-\frac{1}{4} {\bar
X}^IT^{-}_{\mu\nu}+{\rm fermionic\;term}. \ee The last term in the
gaugini transformation is because of the special superconformal
transformation given by the parameters $\eta_i$.

In the ansatz we are considering here, the covariant derivative is
just a simple derivative and since in our ansatz $X^I$ are
constant therefore the first terms is zero. The last term is also
zero because we have  already fixed the conformal gauge. On the
other hand since there is no non-zero fermion in the definition of
${\cal F}^{I-}_{\mu\nu}$, we arrive at \be \delta
\Omega^I_i=\frac{1}{2}\epsilon_{ij} \gamma^\mu\gamma^\nu{\cal
F}_{\mu\nu}^{I-}\epsilon^j \propto A^I \epsilon_{ij}\epsilon^j.
\ee So for $A^I=0$ we would expect to get a supersymmetric
solution, while for $A^I\neq 0$ it would be non-supersymmetric.

As a conclusion we note that the attractor equations are given in
terms of the scalar fields and the functions $A^I$. In order to
find the value of the scalar fields in terms of the black hole
charges one first needs to find $A^I$ in terms of the moduli
fields. This can be done by extremizing the entropy function with
respect to other parameters $v_1,v_2,x^I$ and $\omega$. Doing so,
one finds \bea \frac{\partial {\cal E}}{\partial
x^K}=0&=&4(\frac{1}{v_1}-\frac{1}{v_2})({\bar F}_K- {\bar
x}^IF_{IK})-F_{KIJ}A^IA^J-{\bar \omega}{\bar F}_{KI}{\bar A}^I\cr
&&\cr &+&{\bar \omega}F_{KI}{\bar
A}^I+32\omega{\bar\omega}(-v_1^{-1}-v_2^{-1}+\frac{1}{8}
\omega{\bar
\omega})F_{\hat{A}K}+256(v_1^{-1}-v_2^{-1})^2F_{\hat{A}K},\cr
&&\cr \frac{\partial {\cal E}}{\partial v_1}=0&=&-8v_2^{-1}{\bar
F}_Ix^I+2F_{IJ}A^IA^J+4iv_2^{-1} F_{IJ}p^IA^J+2\omega F_{IJ}
\bar{x}^IA^J\cr &&\cr
&+&2iv_2^{-1}\omega\bar{F}_Ip^I+\omega^2\bar{F}_I{\bar x
}^I+\bar{\omega}^2F+64\omega {\bar \omega}(-v_2^{-1}+\frac{1}{8}
\omega {\bar \omega})F_{\hat{A}}\cr &&\cr
&-&512(v_1^{-2}-v_2^{-2})F_{\hat{A}}-c.c.,\cr &&\cr \frac{\partial
{\cal E}}{\partial v_2}=0&=&8v_1^{-1}{\bar
F}_Ix^I-4iv_2^{-1}F_{IJ}p^IA^J -2iv_2^{-1}\omega {\bar F}_I
p^I-2F_{IJ}A^IA^J\cr &&\cr &-&2\omega {\bar F}_IA^I+{\bar
\omega}^2F+64\omega {\bar \omega}(-v_1^{-1}+\frac{1}{8} \omega
{\bar
\omega})F_{\hat{A}}+512(v_1^{-2}-v_2^{-2})F_{\hat{A}}-c.c.,\cr
&&\cr \frac{\partial {\cal E}}{\partial \omega}=0&=&2\omega
F_{IJ\hat{A}}A^IA^J+\frac{1}{4}F_{IJ}\bar{x}^IA^J-2\bar{\omega}\omega
F_{I\hat{A}}\bar{A}^I-\frac{1}{4}\bar{F}_IA^I\cr &&\cr
&+&8\omega(v_1^{-1}-v_2^{-1})\bar{x}^IF_{I\hat{A}}+8\bar{\omega}(-v_1^{-1}-v_2^{-1}+\frac{1}{8}\omega\bar{\omega})(F_{\hat{A}}-\bar{F}_{\hat{A}})\cr
&& \cr &-&64\omega
F_{\hat{A}\hat{A}}\bigg{(}\omega\bar{\omega}(-v_1^{-1}-v_2^{-1}+\frac{1}{8}\omega\bar{\omega})-8(v_1^{-1}-v_2^{-1})^2\bigg{)}.
 \label{Ex}
 \eea
These equations are enough to find $A^I, v_1$ and $v_2$ in terms
of $x^I$. Then by plugging them  into the attractor equations one
can find the moduli $x^I$ in terms of the electric and magnetic
charges of the black hole as expected from attractor behavior.
Finally due to the entropy function formalism the entropy
associated with the black hole is given by the value of the
entropy function at the extremum \be S_{BH}={\cal E}|_{\rm extremum}.
\ee

\subsection{Explicit example}

To see how these attractor equations work, let us consider a
specific theory with three vector multiplets and a prepotential
\be
F(X^0,X^1,X^2,X^3,\hat{A})=-\frac{X^1X^2X^3}{X^0}-C\hat{A}\frac{X^1}{X^0}\;.
\label{pre} \ee This is the theory known as STU model with the
identification \be
\frac{X^1}{X^0}=iS,\;\;\;\frac{X^2}{X^0}=iT,\;\;\;\frac{X^3}{X^0}=iU.
\ee which describes a subsector of the low energy effective action
for tree level Heterotic string theory on $T^4\times T^2$ or
$K_3\times T^2$. For such a prepotential, the equations of motion
derived from the Lagrangian density are invariant under
$SO(2,2)=SL(2,R)\times SL(2,R)$ T-duality symmetry. If we define
the electric and magnetic charges related to the gauge fields as
\bea q_0&=&Q_4,\;\;\;\;q_1=P_4,\;\;\;\;q_2=Q_1,\;\;\;\;q_3=Q_3,\cr
&&\cr p^0&=&P_2,\;\;\;\;p^1=-Q_2,\;\;p^2=P_3,\;\;\;\;p^3=P_1, \eea
we can easily see that under $SO(2,2)$ duality transformations,
$\vec{Q}$ and $\vec{P}$ behave in a way that $Q^2$, $P^2$ and
$Q.P$ given by \bea
Q^2&=&2(Q_1Q_3+Q_2Q_4),\;\;\;P^2=2(P_1P_3+P_2P_4),\cr &&\cr
Q.P&=&(Q_1P_3+Q_3P_1+Q_2P_4+Q_4P_2), \label{QP} \eea remain
invariant. Due to this symmetry one has the freedom to work in a
frame in which $p^0=0$. Therefore using the attractor equation for
$p^0$ in (\ref{p}), we choose $X^0$ to be real and as a result
$A^0$ will also be real.

We can now proceed to solve the equations of motion for this case.
For the moment we assume $C=0$, or in other words we consider the leading
order term.
Setting  $x^I=y^I+iz^I$ with $z^0=0$, one can
see that the most general solutions of the equations of motion are
given by
\be v_1=v_2=\frac{16}{\omega\bar{\omega}},
\ee
and
\be A^I=0,\label{susy} \ee
or
\be
A^I=-\frac{1}{4}(3y^I+iz^I),\label{nonsusy}
\ee
which correspond
to supersymmetric and non-supersymmetric solutions, respectively.
Note that we have fixed our gauge by choosing
$\omega=\frac{1}{2}$.

Let us consider the supersymmetric solution given by (\ref{susy}).
Plugging this solution into the attractor equations given by
(\ref{q}) and (\ref{p}) we can find the
 value of the moduli at the attractor points as follows
\bea
x^0&=&-\frac{1}{16}Q_2\sqrt{\frac{P^4}{P^2Q^2-(P.Q)^2}},\;\;
\frac{x^1}{x^0}=-\frac{P.Q}{P^2}+i\sqrt{\frac{P^2Q^2-(P.Q)^2}{P^4}},\cr &&\cr
\frac{x^2}{x^0}&=&-\frac{1}{2Q_2P_1}(Q_2P_4+Q_1P_3-P_1Q_3)-i\frac{P_3}{Q_2}\sqrt{\frac{P^2Q^2-(P.Q)^2}{P^4}},\cr
&&\cr
\frac{x^3}{x^0}&=&-\frac{1}{2Q_2P_3}(Q_2P_4-Q_1P_3+P_1Q_3)-i\frac{P_1}{Q_2}\sqrt{\frac{P^2Q^2-(P.Q)^2}{P^4}}.
\eea
This solution is physical for $P^2>0$ and $(Q.P)^2<Q^2P^2$.
%Using the attractor equations for $p^I$ and $q^I$ and also equation (\ref{e}) for $e^I$,
%one can obtain the entropy as %entropy function evaluated at the extremum.
Using the entropy function evaluated at the extremum, the entropy
of these supersymmetric black holes is \be
S_{BH}=\frac{\pi}{2}\sqrt{P^2Q^2-(P.Q)^2}. \ee
Similarly one can
proceed to the other solution which is non-BPS, (\ref{nonsusy}),
to find the values of the moduli at the horizon
 \bea
x^0&=&-\frac{1}{32}Q_2\sqrt{\frac{P^4}{-P^2Q^2+(P.Q)^2}}\;,\;\;
\frac{x^1}{x^0}=-\frac{P.Q}{P^2}+i\sqrt{\frac{-P^2Q^2+(P.Q)^2}{P^4}}\;,\cr
&&\cr
\frac{x^2}{x^0}&=&-\frac{1}{2Q_2P_1}(Q_2P_4+Q_1P_3-P_1Q_3)-i\frac{P_3}{Q_2}
\sqrt{\frac{-P^2Q^2+(P.Q)^2}{P^4}}\;,\cr &&\cr
\frac{x^3}{x^0}&=&-\frac{1}{2Q_2P_3}(Q_2P_4-Q_1P_3+P_1Q_3)-i\frac{P_1}{Q_2}
\sqrt{\frac{-P^2Q^2+(P.Q)^2}{P^4}}\;, \eea which is the same as
supersymmetric case, but with a minus sign in $P^2Q^2-(P.Q)^2$.
This corresponds to the case where $P^2>0$ and $(Q.P)^2>Q^2P^2$.
This is the non-supersymmetric black hole solution with the
entropy \be S_{BH}=\frac{\pi}{2}\sqrt{-P^2Q^2+(P.Q)^2}\;. \ee In
the more simplified BPS case with only 4 charges non-zero which
are given by \be
P_1=P_3=P_0,\;\;\;Q_2=Q_4=-Q_0,\;\;\;Q_1=Q_3=P_2=P_4=0,\ee it can
be seen that the non-BPS solution can be obtained (up to a
normalization in our notation) by canonical transformation on BPS
solution \cite{Kallosh:2006bx} which is
 \be
P_1=P_3=P_0,\;\;\;-Q_2=Q_4=-Q_0,\;\;\;Q_1=Q_3=P_2=P_4=0.
\ee
This canonical transformation in leading order preserves the effective
potential and entropy of the black hole. We
note, however, that it is not obvious if higher order corrections
would respect this canonical transformation.

The next step is to consider higher order corrections which
correspond to the case where $C\neq 0$. To do this one needs to
solve the equations with $C\neq 0$. Since these equations are
algebraic equations in principle one can solve them. In particular
for the supersymmetric case doing so, in the gauge of
$\omega=\frac{1}{2}$, one finds \cite{Sahoo:2006rp}

\bea
x^0&=&-\frac{1}{16}Q_2\sqrt{\frac{P^2(P^2+512C)}{P^2Q^2-(P.Q)^2}}\;,\;\;
\frac{x^1}{x^0}=-\frac{P.Q}{P^2}+i\sqrt{\frac{P^2Q^2-(P.Q)^2}{P^2(P^2+512C)}}\;,\cr
&&\cr
\frac{x^2}{x^0}&=&-\frac{1}{2Q_2P_1}(Q_2P_4+Q_1P_3-P_1Q_3)-i\frac{P_3}{Q_2}
\sqrt{\frac{P^2Q^2-(P.Q)^2}{P^2(P^2+512C)}}\;,\cr &&\cr
\frac{x^3}{x^0}&=&-\frac{1}{2Q_2P_3}(Q_2P_4-Q_1P_3+P_1Q_3)-
i\frac{P_1}{Q_2}\sqrt{\frac{P^2Q^2-(P.Q)^2}{P^2(P^2+512C)}}\;,
 \eea
with $v_1=v_2=64$ and $A^I=0$. The entropy is given by \be S_{\rm
BH}=\frac{\pi}{2}\sqrt{P^2Q^2-(P.Q)^2}\;\sqrt{1+\frac{512C}{P^2}}\;.
\ee To compare this with the results given in terms of the special
geometry \cite{{LopesCardoso:1998wt},{Mohaupt:2005jd}}
 it is instructive
to rewrite the corrected entropy in terms of the prepotential.
From the entropy function one gets \be S_{\rm
BH}=2\pi(-\frac{1}{2}q_Ie^I-16i(\omega^{-2}F-\bar{\omega}^{-2}\bar{F})).
\ee By making use of the attractor equations
$q_I=4i(\bar{\omega}^{-1}\bar{F}_I-{\omega}^{-1}{F}_I)$ and
$e^I=4(\bar{\omega}^{-1}\bar{x}^I+{\omega}^{-1}{x}^I)$ the entropy
reads \be S_{\rm BH}=\pi\left(8(p^IF_I-q_Ix^I)-256\;{\rm
Im}\;(F_{{\hat A}})\right), \ee which is the same as that obtained
in \cite{{LopesCardoso:1998wt},{Mohaupt:2005jd}}.

\section{Black hole partition function}

In the previous section we have studied the generalized attractor
equations in presence of higher order corrections using entropy
function formalism. The aim of this section is to use this
formalism to understand the physical interpretation of these
equations better. To do this we compare the entropy function
formalism to \cite{Ooguri:2004zv}. In the following we shall
review the relevant part of the paper.

The attractor equations for BPS black hole in ${\cal N}=2$ four
dimensional supergravity could be used to express the entropy of
the extremal black holes as the Legendre transformation
of a function, ${\cal F}$, which is given by
the imaginary part of the prepotential evaluated at the attractor
point
\be
S_{\rm BH}(\vec{p},\vec{q})={\cal
F}(\vec{e},\vec{p})-e^I \frac{\partial{\cal
F}(\vec{e},\vec{p})}{\partial e^I}\equiv {\cal F}(\vec{e},\vec{p})-e^Iq_I\;,
\label{eee}
\ee
where $(\vec{q},\vec{p})$ are electric and magnetic charges of the black
hole and $e$ is the electric potential defined by
$e^I=-\frac{\partial S_{\rm BH}}{\partial q_I}$. It is natural to define
a mixed partition function for black hole as follows
\be
Z_{\rm BH}(\vec{e},\vec{p})=\sum_{\vec{q}}d(\vec{p},\vec{q})e^{
\vec{e}.\vec{q}}.
 \label{mic}
\ee
Here $d(\vec{p},\vec{q})$ is integer black hole degeneracy and $\ln d(\vec{p},
\vec{q})$ is the
microcanonical entropy. Therefore one leads to the following
expression for the black hole partition function in terms of
the function
${\cal F}$ \footnote{Since it is a mixed partition function,
{\it a priori} it is not clear whether one can interpret ${\cal
F}$ as the black hole free energy. We would like to thank
C. Vafa for a discussion on this point.}
\be Z_{\rm BH}(\vec{p},\vec{e})=e^{{\cal
F}(\vec{p},\vec{e})}. \ee

Since in the entropy function formalism one gets the attractor
equations from the equations of motion, it is natural to ask
if we can follow the above procedure to define a (mixed) partition function
using the entropy function formalism.

Form entropy function formalism we learned that the entropy
of the extremal black hole is given by
\be
S_{\rm BH}(p^I,q_I)=2\pi\left(e^I\frac{\partial f}{\partial
e^I}-f\right)=2\pi(e^Iq_I-f). \label{waldf}
\ee

Comparing the Wald formula in the form of (\ref{waldf}) and the fact
that the equations of motion would lead to the attractor equations, it is
tempting to follow OSV proposal to define a partition function
for the corresponding extremal black hole as $Z_{\rm BH}=e^{-f}$.

Let us apply the above procedure to a toy model given by the
following action \be S=\frac{1}{16\pi}\int
d^4x\;\sqrt{-G}\;(R-F^2). \ee Consider an extremal black hole
solution with near horizon geometry given by \be
ds^2=v_1(-r^2dt^2+\frac{dr^2}{r^2})+v_2(d\theta^2+\sin^2\theta
d\phi^2)\;,\;\;\;\;F_{rt}=e\;. \ee So that
$f(v_1,v_2,e)=\frac{1}{2}v_1v_2(\frac{v_1-v_2}{v_1v_2}
+\frac{e^2}{v_1^2}).$  Extremizing $f$ with respect to $v_1,v_2$
we get $v_1=v_2=e^2$. On the other hand we have $q=\frac{\partial
f}{\partial e}=e$ and therefore from (\ref{waldf}) we find $S_{\rm
BH}=\pi q^2$.

Alternatively, using the fact that $f=e^2/2$  one can define a partition function as $Z(e)=e^{-e^2/2}$ and
therefore we find the
microscopic degrees of freedom as follows
\be d(q)=\int de\;
e^{2\pi(qe-\frac{1}{2}e^2)}=e^{\pi q^2}
\;\;\;\;\Rightarrow\;\;\;S_{\rm micro}=\ln d(q)=\pi q^2\;,
\ee
in agreement with the black hole entropy.
Therefore we can conclude
that the entropy of the black hole is exactly given by the microcanonical entropy.
It is also possible to consider higher order corrections to the
action. In the present case the corrections can be given by the Gauss-Bonnet
action which leads to a correction as $d(q)=e^{\pi q^2+2\pi \lambda}$ or $S_{\rm micro}= \pi q^2+2\pi \lambda$, in agreement with the Wald formula
for the black hole  entropy in the presence of the Gauss-Bonnet
term \cite{Alishahiha:2006ke}.
Here $\lambda$ is the coefficient of the Gauss-Bonnet term.

To be more realistic we consider the system we have studied in the
previous section. By making use of the results in the previous
section one finds \bea f(v_1,v_2,e^I,p^I,x^I,\omega)&=&\frac{1}{2}
v_1v_2\bigg{\{}i({v_1}^{-1}-{v_2}^{-1})x^I{\bar F}_I
-\frac{i}{4}F_{IJ}A^IA^J -\frac{i}{4}\omega {\bar F}_{I}A^I \cr
&&\cr & +&\left( 8i\omega{\bar
\omega}(-v_1^{-1}-v_2^{-1}+\frac{1}{8} \omega{\bar
\omega})+64i(v_1^{-1}-v_2^{-1})^2\right)F_{\hat{A}} \cr &&\cr &+&
\frac{i}{8}{\bar \omega}^2 F+c.c. \bigg{\}}. \label{hhh} \eea In
the supersymmetric case from equations (\ref{Ex}) we find $A^I=0$
and $v_1=v_2=\frac{16}{\omega\bar{\omega}}$. So one arrives at \be
f(e^I,p^I)=-2{\rm{Im}}\left(\left(\frac{4}{\omega}\right)^{2}F(x^I,\omega)\right),
\ee where $x^I=\frac{\omega}{8}(e^I+ip^I)$. It is worth noting
that in comparison with \cite{Ooguri:2004zv} $f$ can be identified
with $-{\cal F}$ and therefore following \cite{Ooguri:2004zv} the
system can be described as a mixed ensemble.

Since, in general, the equations
(\ref{Ex}) have two solutions, supersymmetric and non-supersymmetric,
one might naturally expect that the situation
would also go through the non-supersymmetric case.
In fact from the equations of motion one may first obtain $A^I$ and
therefore using the relation
$\bar{A}^I+\frac{\bar{\omega}}{2}x^I=v_1^{-1}(e^I+iv_1v_2^{-1}p^I)$
we can find the moduli in terms of $e^I$ and $p^I$. Plugging the
results into (\ref{hhh}) one finds $f$ as a function of  $e^I$ and
$p^I$ and thereby the partition function can be evaluated along the
supersymmetric case. Partition function for non-BPS solution has also
been studied in \cite{{Parvizi:2006uz},{Ferrara:2006em}}.

%As an aside it is interesting to compare the entropy function of the
%BPS and non-BPS solutions in this theory. For non-BPS solution in
%the leading order one finds ${\cal{E}_{\rm non-BPS}}=
%\frac{2048\pi}{y_0}z_1z_2z_3$, whereas for BPS we get
%${\cal{E}_{\rm BPS}}=\frac{512\pi}{y_0}z_1z_2z_3$. Therefore we observe
%${\cal E}_{\rm non-BPS}=4{\cal E}_{\rm BPS}$. The factor of four
%difference between BPS and non-BPS has also been observed in
%\cite{{Parvizi:2006uz},{Ferrara:2006em}}. We note, however, that this is
%just the leading order effect and going beyond the tree level, in general,
%we won't get such a simple relation.

Regarding the fact that the entropy function formalism could
simply reproduce the known results for BPS case and also is
powerful enough to be generalized to non-BPS solution, one then
might naively think that we can generalize it for theories
without supersymmetry as well. We note, however, that it is not
obvious whether this is going to be the case.
In fact {\it a priori}  it is not clear if the black hole could be
described by a mixed ensemble even though the entropy is given by a
Legendre transformation of the function $f$.
We will come back to this point in the next section.

\section{Discussions}

In this paper we have shown how the entropy function can reproduce
the generalized attractor equations and also how to generalize them
while higher order corrections are also taken into account.
In fact in the supersymmetric model we have studied in this paper
we showed that one of the attractor equations comes out as
the equation of motion and the other comes as the supersymmetry
condition.

Having had the attractor equations in general form we have also
tried to see if this can help us to define a partition function
for the extremal black hole following \cite{Ooguri:2004zv}. In
particular we have considered a particular example which is an
extremal black hole in four dimensional supergravity obtained by
compactification of type IIA on Calabi-Yau 3-fold. In this example
we could identify the function $f$ with $-{\cal F}$ and thereby to
define a mixed partition function as $e^{-f}$.

We note, however, that these black holes can also be studied from
Heterotic string point of view. It is then natural to ask if this
description also leads to the same conclusion. Therefore it is
worth to reconsider the model from the Heterotic string point of
view. To do this we shall restrict ourselves to two charged black
hole, though the generalization for higher charges is
straightforward.

For two charged black hole it is known that in leading order the
entropy is zero while higher order corrections (Gauss-Bonnet in
the present case) stretches the horizon leading to non-zero
entropy. Taking into account the Gauss-Bonnet term and doing  the
same computations as that in the previous section (see also
\cite{Sen:2005iz}) we find $f=0$. Therefore it seems impossible to
use the OSV procedure to define a (mixed) partition function. In
fact there is a proposal for the statistical ensemble one may
associate to this black hole due to Sen \cite{Sen:2004dp} who
conjectured that, after taking into account the holomorphic
anomaly, a grand canonical ensemble underlies the system. On the
other hand from our considerations in the previous section which
has been done in the type IIA dual description we have been able
to get a mixed partition function using entropy function mechanism
(see also \cite{Dabholkar:2004yr} where the author has considered
the system as a mixed ensemble and confirmed OSV conjecture
without taking into account the holomorphic anomaly term.).

Therefore we face a puzzle, namely, studying the system from two
different point of views, leads to two different ensembles. So far
we do not have a good interpretation of this observation. As far
as the technical point is concerned we note that changing type IIA
description to Heterotic description we have lost one of the
attractor equations. As we have mentioned one of the equations
comes from the equations of motion while the other one is the
condition we get from supersymmetric condition which as we have seen,
depends on the way we incorporate the supersymmetry in the theory.

We also note that it might be related to the fact that OSV
proposal is not symplectic invariant. In particular it begins with
a symplectic invariant answer ( Wald formula)  and performs a
Legendre transform and an inverse Laplace transform both of which
are non-symplectic invariant operation. It is not guaranteed that
the final answer will be symplectic invariant.  In fact the
entropy function seems to be a more natural object than the
Lagrangian density $f$, (the latter is not symplectic invariant
while the former is) \footnote{ We would like to thank A. Sen for
a discussion on this point.}.

As a final remark we note that the basic point in entropy function
formalism is the fact that near horizon field configuration of
these extremal black holes is fixed just by using the symmetries
of near horizon geometry that is $AdS_2\times S^{2}$.
%Therefore the values of the scalar fields at
%the horizon are fixed and solving equations of motion, obtained
%from Lagrangian of the theory, shows that their values are just
%given by the black hole electric and magnetic charges.
This is, in fact, the notion of attractor mechanism which means
that the value of the scalar fields at the horizon are independent
of their values at infinity and they are fixed by the black hole
charges. Moreover the entropy of the black hole is just given by
the black hole charges too. Therefore one may conclude that the
entropy function formalism leads us to the fact that the near
horizon field configuration has enough information about the
corresponding extremal black hole. In this sense, in comparison
with AdS/CFT
\cite{{Maldacena:1997re},{Gubser:1998bc},{Witten:1998qj}}, one
might suspect that the attractor behavior plays the role of the
decoupling limit in this context.

It is worth noting that the black hole attractor mechanism can also be
treated as the holographically dual to a conformally invariant quantum
mechanics \cite{Gaiotto:2004ij}. This might also indicate that the
near horizon modes have enough information about the whole system.
It would be very interesting to understand this connection better.
To do this, it might be useful to consider the supersymmetric case
where one may use the results of \cite{{Ooguri:2005vr},{Gukov:2005bg}}.
In this supersymmetric case where we consider type IIA string theory
compactified on Calabi-Yau 3-fold the four dimensional theory
may have an extremal black hole solution which can be studied using the
$AdS_2\times S^2$ background. In this case the flux data on the
$AdS_2\times S^2$ geometry is mapped to the charges of the dual black hole
and its entropy is the logarithm of the norm of the Hartle-Hawking wave
function on $AdS_2\times S^2$ \cite{{Ooguri:2005vr},{Gukov:2005bg}}.

\noindent\textbf{Acknowledgments}

We would like to thank Ashoke Sen, Alireza Tavanfar and Cumrun
Vafa for useful  discussions and comments.

\end{document}